%% file: main.tex
\documentclass[graybox]{svmult}

\usepackage{type1cm}        % activate if the above 3 fonts are
\usepackage{makeidx}         % allows index generation
\usepackage{graphicx}        % standard LaTeX graphics tool
\usepackage{multicol}        % used for the two-column index
\usepackage[bottom]{footmisc}% places footnotes at page bottom

\usepackage{newtxtext}       % 
\usepackage[varvw]{newtxmath}       % selects Times Roman as basic font
\usepackage{hyperref}
\usepackage{xcolor}

\hypersetup{
  colorlinks   = true, 
  urlcolor     = blue,
  linkcolor    = blue, 
  citecolor   = blue 
}
\makeindex             % used for the subject index
\begin{document}

\title*{Extreme magnetic fields around black holes}
\author{Koushik Chatterjee\orcidID{0000-0002-2825-3590}, \\Matthew Liska\orcidID{0000-0003-4475-9345}, \\Alexander Tchekhovskoy\orcidID{0000-0002-9182-2047}, \\Sera Markoff\orcidID{0000-0001-9564-0876}, and \\Ramesh Narayan\orcidID{0000-0002-1919-2730}}
\authorrunning{Chatterjee et al.}
\institute{Koushik Chatterjee \at Institute for Research in Electronics and Applied Physics, University of Maryland, 8279 Paint Branch Drive,
College Park, MD 20742, USA, \email{kchatt@umd.edu}
\and Matthew Liska \at Center for Relativistic Astrophysics, Georgia Institute of Technology, Howey Physics Bldg, 837 State St NW, Atlanta, GA 30332, USA, \email{mliska3@gatech.edu}
\and Alexander Tchekhovskoy \at Center for Interdisciplinary Exploration \& Research in Astrophysics (CIERA), Physics \& Astronomy, Northwestern University, Evanston, IL 60202, USA, \email{atchekho@northwestern.edu}
\and Sera Markoff \at Anton Pannekoek Institute for Astronomy, University of Amsterdam, Science Park 904, 1098 XH Amsterdam, The Netherlands, \email{S.B.Markoff@uva.nl}
\and Ramesh Narayan \at Black Hole Initiative at Harvard University, 20 Garden Street, Cambridge, MA 02138, USA, \email{rnarayan@cfa.harvard.edu}
}
\maketitle

\abstract{Recent results of the event horizon-scale images of M87* and Sagittarius A* from the Event Horizon Telescope Collaboration show that strong magnetic fields are likely present around the central black holes (BHs) in these sources. Magnetically arrested disks (MADs), the end stage of magnetic flux saturation around BHs, are especially rich in horizon-scale physics due to the presence of powerful jets and magnetic flux eruptions that provide significant feedback on the accretion mechanism. Here, we present an overview of our current knowledge about the magnetic field evolution in numerical simulations of accreting BHs, focusing on jet launching, black hole-interstellar medium feedback, and black hole imaging of MADs. We find that misaligned MAD accretion flows seemingly exhibit jet ejection cycles that could produce flaring states in radio-quiet active galactic nuclei. Further, we show that advances in horizon-scale interferometric telescopes could identify disk misalignment by imaging the disk-jet connection region.
}

\section{Introduction}
\label{sec:intro}

\begin{figure}[t]
\begin{center}
    
    \includegraphics[height=1.75in,trim=0cm 0cm 0cm 0cm, clip]{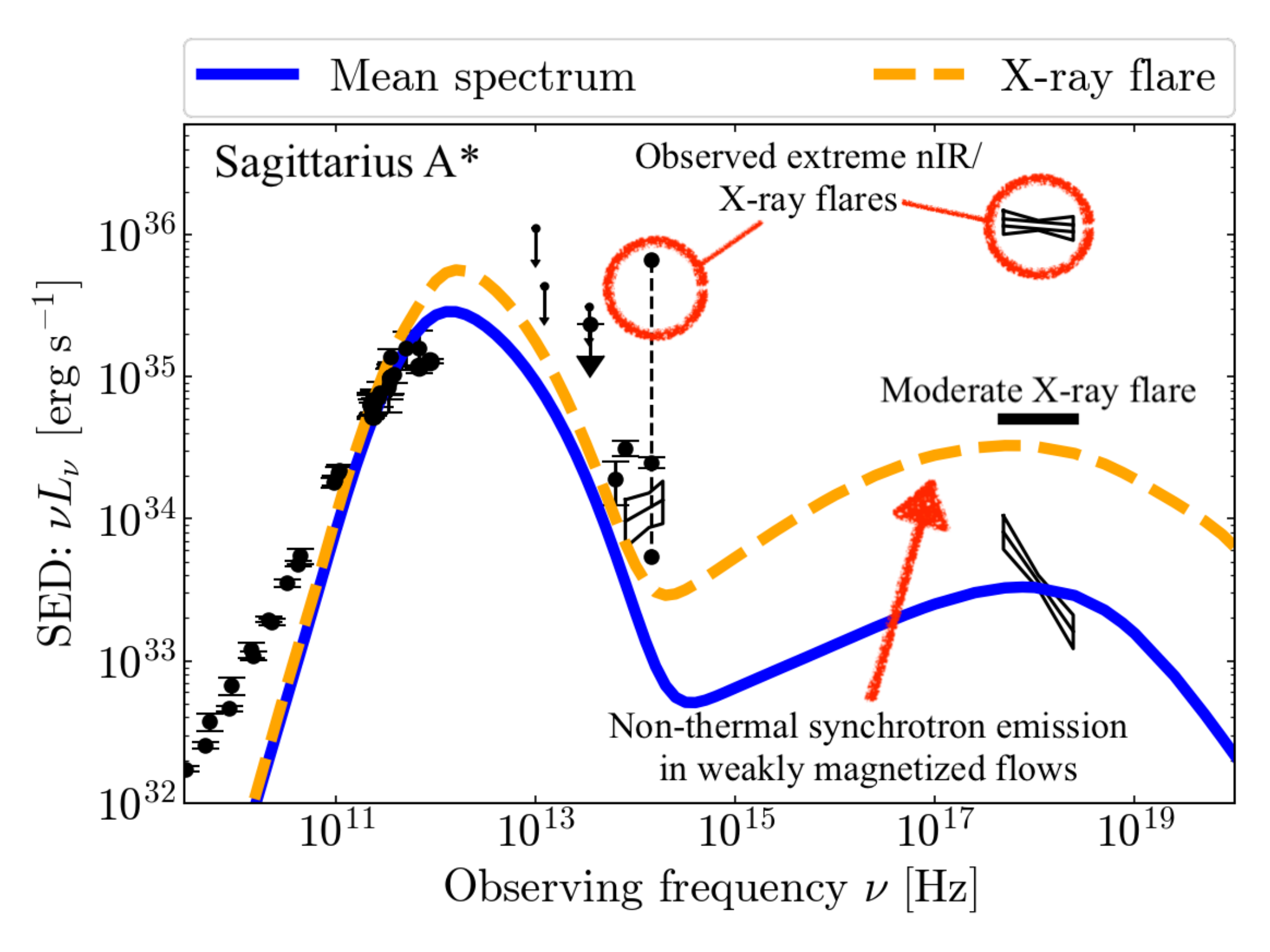}
    \includegraphics[height=1.67in,trim=0cm 0cm 0cm 0cm, clip]{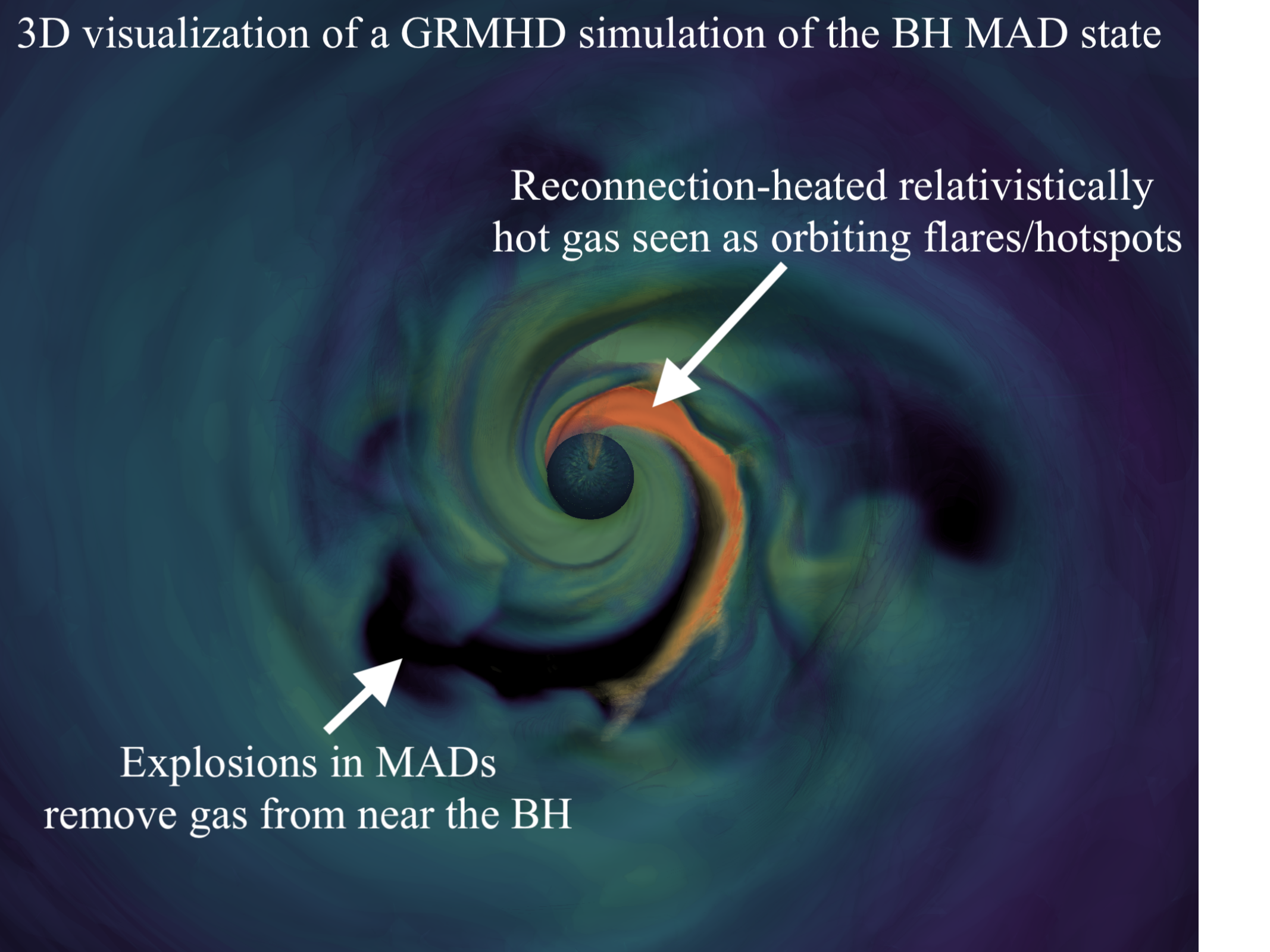}
	\caption{(\textit{Left}) Magnetised gas turbulence triggers particle acceleration that leads to simultaneous moderate nIR/X-ray flares, but cannot explain the most energetic flares \cite{Chatterjee2021}. We show a multi-wavelength SED (solid and dashed lines) of Sagittarius A* from a GRMHD simulation of a weakly magnetised accretion flow around a spinning BH assuming synchrotron emission from both thermal and non-thermal (i.e. accelerated) electron populations. Sgr A* data is shown in black (see \cite{Chatterjee2021} for more details). Note that we ignore synchrotron self-Compton calculations in this SED. (\textit{Right}) Magnetically arrested disks (MADs) exhibit magnetic explosions that heat up gas (red) in the disk (green/blue), a mechanism favourable for producing extreme high-energy flares.}
    \label{fig:flares}
\end{center}
\end{figure}

While the idea of magnetically dominated accretion flows has been around for at least 50 years \cite{BK1974}, this mode of accretion, dubbed magnetically arrested disks (MADs), truly became a subject of interest over the past 2 decades or so \cite{Igumenshchev2003, Narayan2003}. Part of the interest stemmed from advances in computational development of 3D general relativistic magnetohydrodynamic (GRMHD) codes that enabled self-consistent simulations of black hole (BH) accretion. On the observational front, the multitude of Sagittarius A* (or Sgr A*) X-ray flares captured by the Chandra observatory \cite{Baganoff2001, Nowak2012, Haggard2019} provided indications of particle acceleration, possibly driven by reconnection of magnetic fieldlines close to the BH \cite{Markoff2001, Yuan2003}. Figure~\ref{fig:flares} shows that our standard picture of magnetorotational instability (MRI)-driven accretion produces X-ray and near-infrared (nIR) flares too weak to explain some of Sgr A*'s brightest flares \cite{Do2019, Haggard2019}, suggesting the need for stronger magnetic fields. More recently, the polarised horizon-scale EHT images of M87 and Sgr A* also favour strong vertical magnetic fields \cite{EHTpaperVIIIM872021, EHTpaperVIIISgrA2024}. Indeed, the MAD state could be a natural outcome of starved BHs (i.e., accreting at highly sub-Eddington rates), which is the case for both Sgr A* and M87. There is seemingly an abundant supply of poloidal magnetic field either by advecting $\mu$Gauss-level magnetic fields from the interstellar medium (ISM) \cite{Narayan2003, Haverkorn2013} or from magnetised stellar winds (as thought to occur in Sgr A*; \cite{Ressler2020}), or by in-situ generation of poloidal flux via dynamo processes in the disk \cite{Liska2020}, favouring the emergence of the MAD state. Motivated by these advances in theory, numerics and observations, we discuss the many features of MAD accretion flows and their outflows, with a focus on recent results of misaligned MAD flows from Chatterjee et al. \cite{Chatterjee2023tiltedMADs}.

\section{Revisiting magnetically arrested accretion flows}
\label{sec:MADs}

In this section, we introduce some salient features of the MAD state. We divide our discussion into two parts: time-averaged and time-dependent disk/jet properties. The former category of analysis allows us to understand broad features of MADs and cross-compare between multiple simulations, while the latter involves tracking the gas and magnetic field dynamics over time with the aim of studying the evolution of magnetic explosions often seen in MADs, and their effect on jet morphology. For our discussion, we make use of results from MAD simulations, performed using the GRMHD code \texttt{H-AMR} \cite{Liska2022}, from a number of published papers \cite{Chatterjee2020, Chatterjee2022, Chatterjee2023tiltedMADs}. These results are largely consistent with simulations from other codes. 

We focus on three quantities: the accretion rate $\dot{M}=-\iint \rho u^r dA_{\theta\varphi}$, the poloidal magnetic flux $\Phi_{\rm BH}=(1/2)\iint |B^r| dA_{\theta\varphi}$ calculated at the event horizon, and the total outflow power $P=\dot{M}c^2-\dot{E}$. For these quantities, we require the gas density $\rho$, the radial component of the gas 4-velocity $u^r$, the radial component of the 3-magnetic field vector $B^r$ and the total energy flux in the radial direction $\dot{E}=\iint T^r_{t} dA_{\theta\varphi}$, defined using the stress-energy tensor $T^{\mu}_{\nu}$ and the area element $dA_{\theta\varphi} = \sqrt{-det(g_{\mu\nu})}d\theta d\varphi$. We assume natural units $G=c=1$, which implies that our characteristic length-scale, the gravitational radius $r_{\rm g}=GM/c^2$, and the corresponding light-crossing time-scale $t_{\rm g}=r_{\rm g}/c$, are normalised to the BH mass $M$. For our simulations, we can set $M=1$. Further, we can normalise the magnetic flux and the outflow power with the time-averaged accretion rate $\langle\dot{M}\rangle$, giving us the dimensionless magnetic flux $\phi_{\rm BH}=\Phi_{\rm BH}/\sqrt{\langle\dot{M}\rangle}$ and the outflow efficiency $\eta=P/\langle\dot{M}\rangle$ (note that $r_{\rm g}$ and the speed of light $c$ are unity according to our units). We refer the reader to Porth et al. \cite{Porth2019} for more details about these quantities.

\begin{figure}[t]
    \centering

    \includegraphics[width=0.49\columnwidth,trim= 0pt 0pt 0pt 60pt, clip]{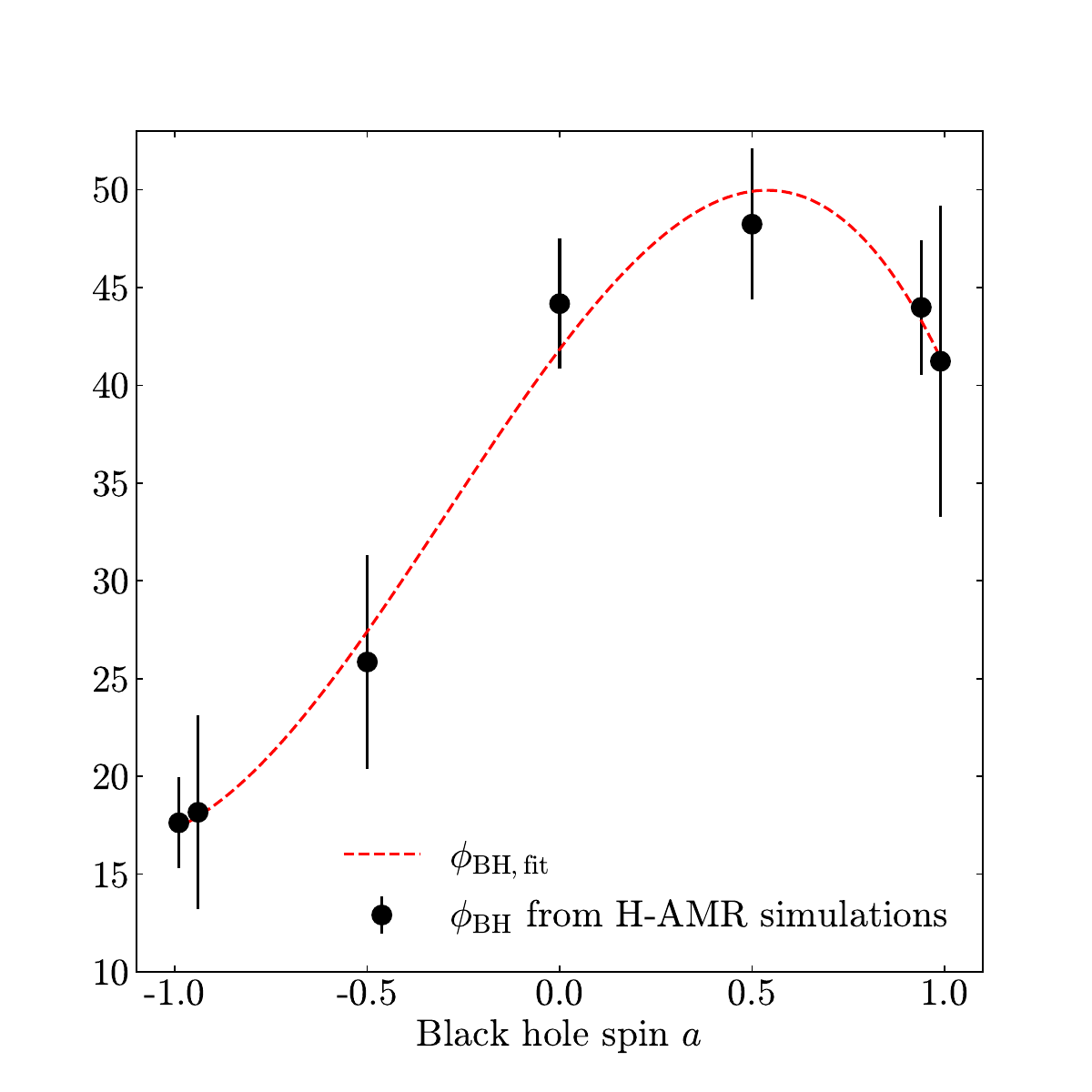}
    \includegraphics[width=0.49\columnwidth,trim= 0pt 0pt 0pt 60pt, clip]{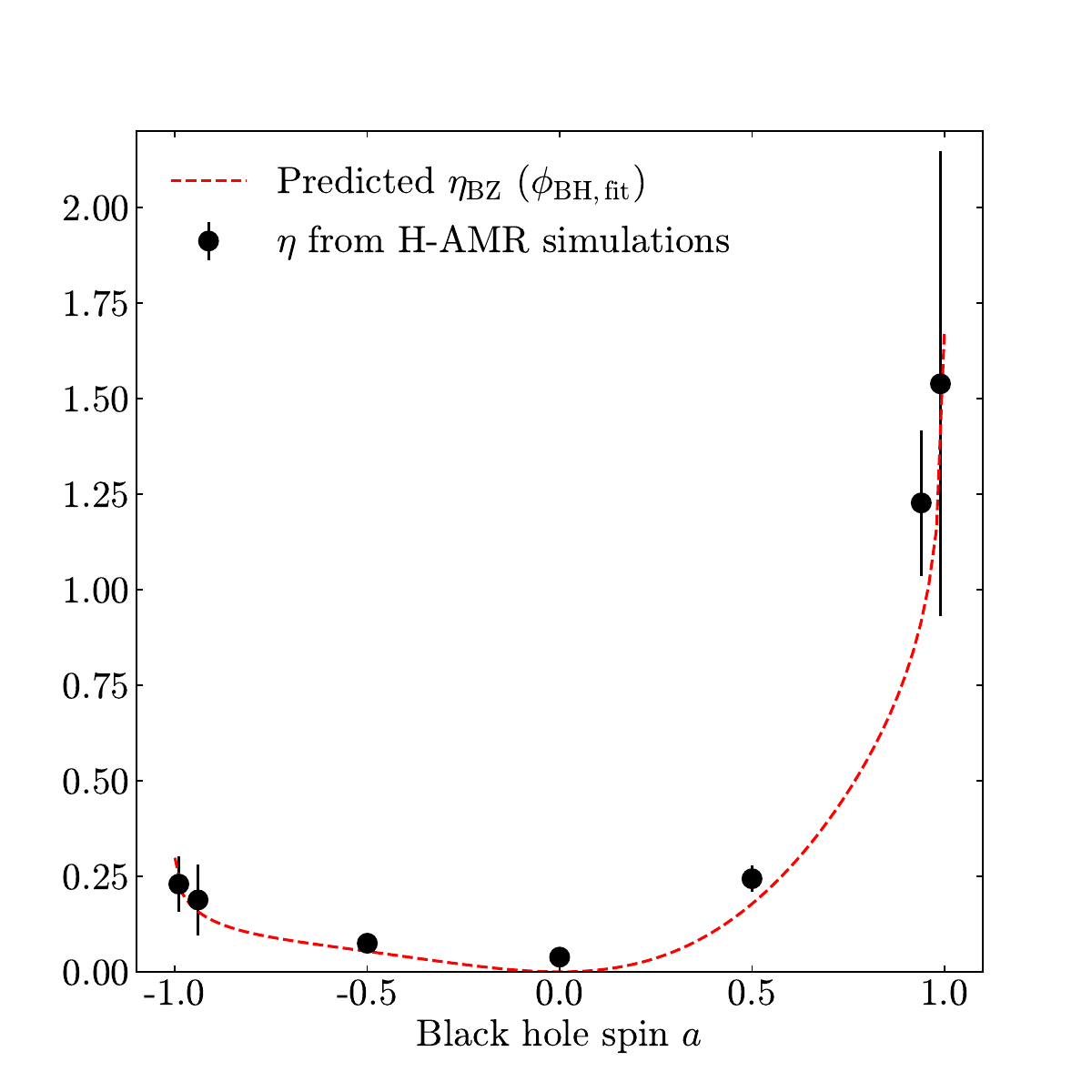}
    \caption{(\textit{Left}) We show the accretion-rate normalised magnetic flux at the BH horizon for GRMHD simulations of MADs around a variety of BH spins \cite{Chatterjee2022}. We also show the $1\sigma$ standard deviation for each model to indicate the amount of variability in $\phi_{\rm BH}$. The dashed line shows a cubic fit to the numerical data, using the \texttt{scipy} function \texttt{curvefit}.  (\textit{Right}) We show the corresponding outflow efficiency $\eta$ along with the $1\sigma$ standard deviation for each model. The dashed line shows the expected Blandford-Znajek (BZ) jet efficiency $\eta_{\rm BZ} \propto \phi_{\rm BH, fit}^2\Omega_{\rm H}^2$ \cite{BZ1977, Tchekhovskoy2012, Chatterjee2023JP}, where $\Omega_{\rm H}=a/2r_{\rm H}$ is the horizon angular frequency and $r_{\rm H}=(1+\sqrt{1-a^2})r_{\rm g}$ is the event horizon radius. We find that for BHs with a large prograde spin, (1) the BZ power matches the simulation results as jets dominate the total outflow power, and (2) both $\phi_{\rm BH}$ and $\eta$ exhibit the largest variability among the models, presumably due to powerful and frequent magnetic flux eruptions.
    }
    \label{fig:eta}
\end{figure}

\subsection{Time-averaged properties of MADs: magnetic flux saturation and relativistic jets}
\label{sec:time_avg}

MADs deal with magnetically-saturated BH environments where forces due to magnetic fields are roughly in radial balance with the ram pressure from the accreting gas. It thus follows that such a system should exhibit the maximum horizon-scale poloidal magnetic flux possible given sufficiently large supply from the disk. Any amount of extra magnetic flux accumulated near the horizon results in expulsion of magnetic fluxtubes, or ``magnetic flux eruptions'', creating temporal oscillations of the (largely radial) magnetic flux $\phi_{\rm BH}$ at the BH horizon. We will discuss these eruptions in the next subsection. Figure~\ref{fig:eta} shows the time-averaged $\phi_{\rm BH}$ as a function of BH spin, calculated over $5000\,t_{\rm g}$ using \texttt{H-AMR} simulations \cite{EHTpaperVSgrA2022, Chatterjee2022}. We see that $\phi_{\rm BH}$ has a cubic dependence of the BH spin parameter $a$: $\phi_{\rm BH, fit}= 41.84 + 26.06a -12.68a^2 -14.23a^3$, which is roughly consistent with previous work using GRMHD simulations \cite{Tchekhovskoy2012, Narayan2022, Chatterjee2023tiltedMADs}. While this cubic dependence on spin has not yet been worked out from first principles, it is reasonable to assume that the BH spin should strongly affect the MAD magnetic flux since (1) the magnitude of $\phi_{\rm BH}$ depends on the horizon area (which, in turn, is spin-dependent), and (2) the accretion disk close to the BH is subject to BH frame-dragging. For example, Narayan et al. \cite{Narayan2022} showed that the disk scale-height appears to be closely tied to the BH spin. Additionally, for BHs accreting at or close to their Eddington rate, radiation physics would also strongly affect disk structure \cite{Ricarte2023, Liska2024, Lowell2024}. 

In MADs, the strong magnetic fields near the BH not only modulate disk dynamics, but also power highly efficient relativistic jets. GRMHD simulations show that, in most cases, there is always an outflow of energy (and sometimes mass) in the form of jets and disk winds, provided that the BH environment is sparse enough to allow jet propagation. Figure~\ref{fig:eta} shows one of the most interesting properties of MADs: rapidly spinning prograde BHs can launch jets with efficiencies $\eta$ exceeding 100\% in the MAD state, first shown by Tchekhovskoy et al. \cite{Tchekhovskoy2011}, which provides direct evidence of spin energy extraction from BHs, a direct consequence of the Blandford-Znajek (BZ) mechanism \cite{BZ1977}. Indeed, it has been shown that jets from MADs spindown their BHs, with the equilibrium BH spin close to zero \cite{Tchekhovskoy2012, Chatterjee2022, Narayan2022, Ricarte2023, Lowell2024}, which could have exciting implications for spindown of BHs accreting at super-Eddington rates, e.g., in gamma-ray bursts and ultra-luminous X-ray binaries \cite{Narayan2017}.

\subsection{How MADs evolve over time: the role of magnetic flux eruptions}
\label{sec:time}

\begin{figure}[t]
\centering
\includegraphics[width=0.51\columnwidth,trim= 18pt 0pt 200pt 0pt, clip]{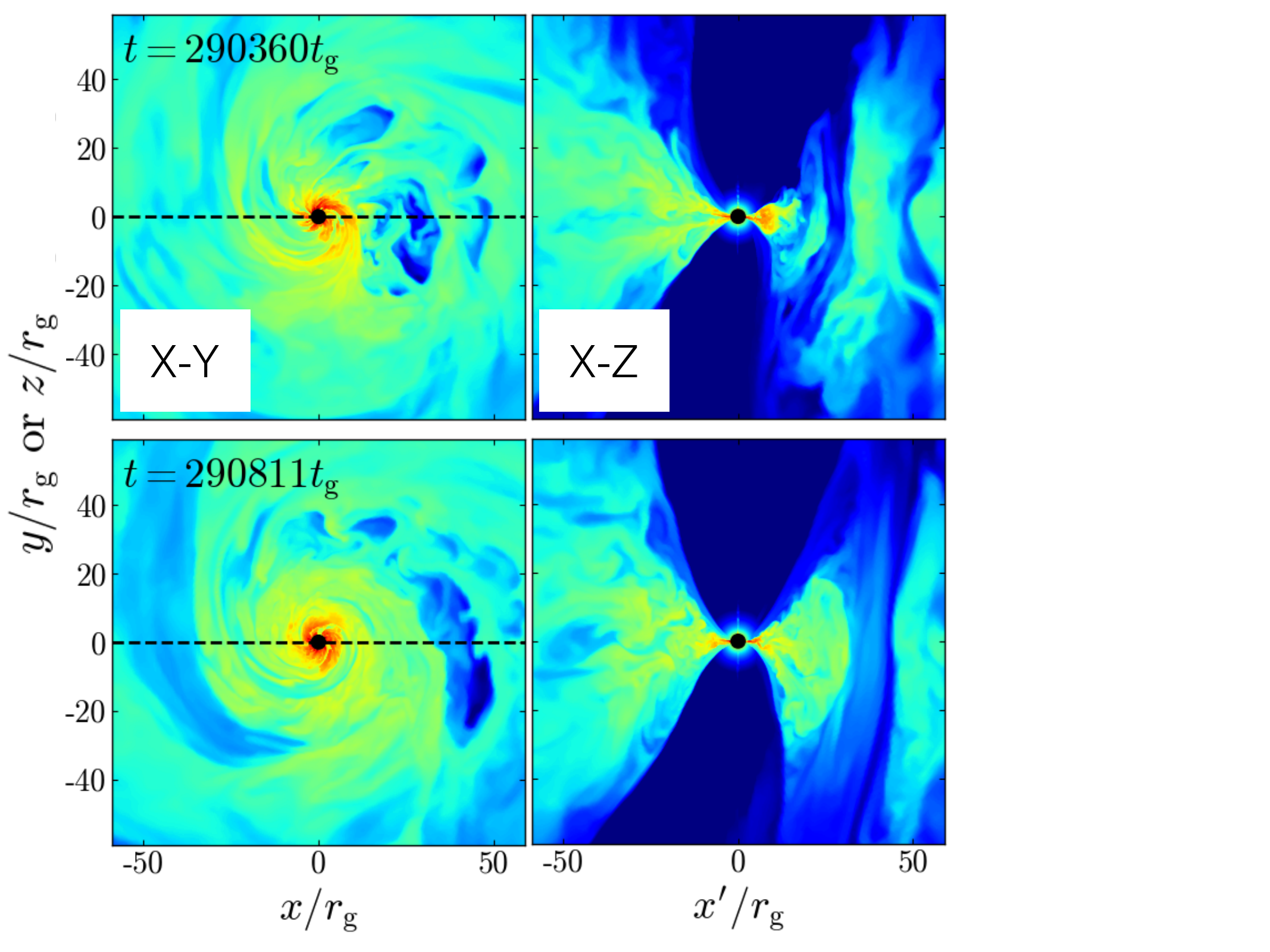}
\includegraphics[width=0.45\columnwidth,trim= 50pt 5pt 50pt 50pt, clip]{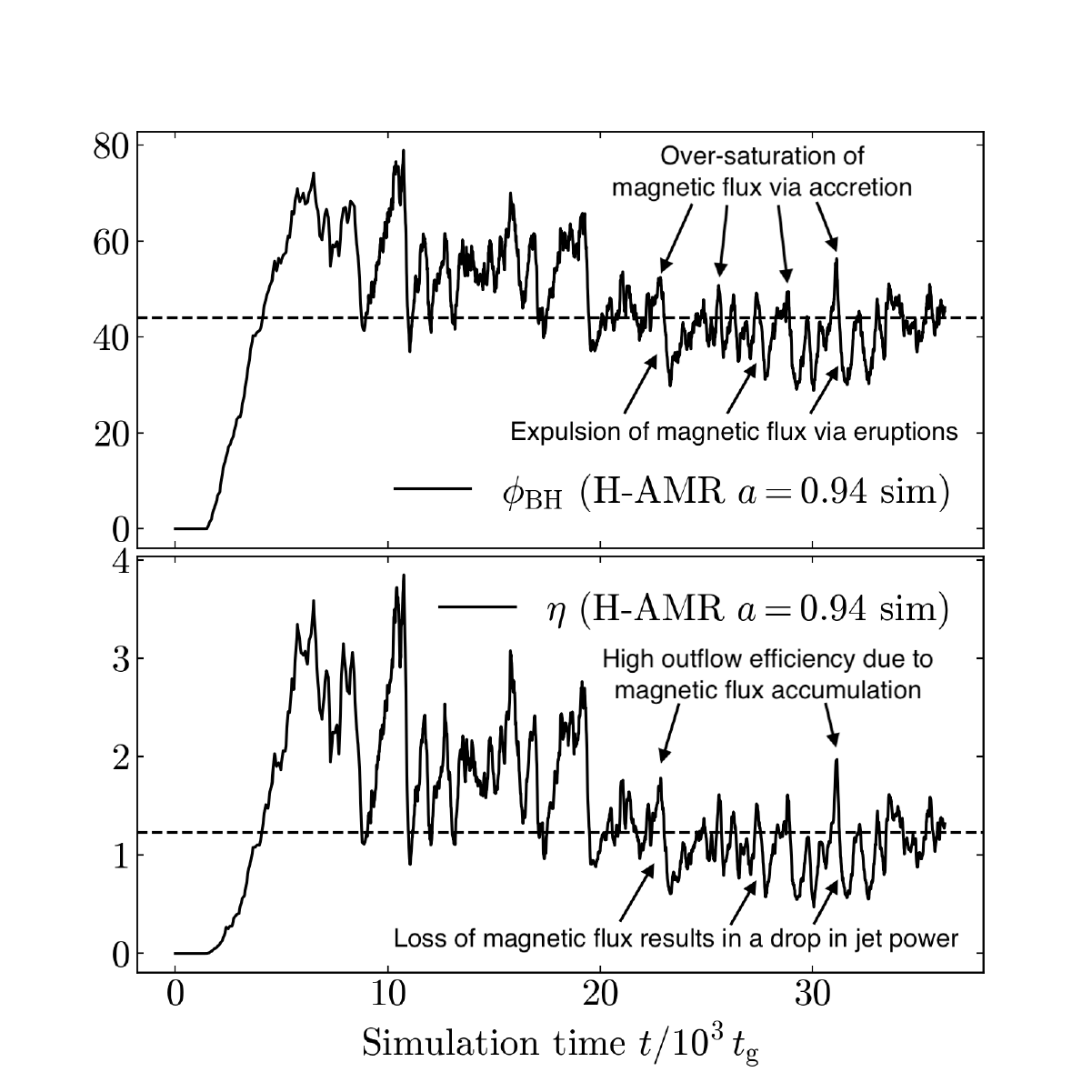}
\caption{(\textit{Left}) We show the x-y and x-z planes of two representative snapshots of a MAD flow around a Schwarzschild BH from Chatterjee et al. \cite{Chatterjee2022}, tracking the movement of a magnetic fluxtube as it propagates into the disk. From the figures, we see that a bundle of fieldlines get separated from the jet and form a fluxtube. The accretion flow gradually shears out the fluxtube over time, giving rise to Rayleigh-Taylor-like (RT) and Kelvin-Helmholtz (KH) instabilities at the fluxtube boundary that could explain nIR flares seen in Sgr A* \cite{Zhdankin2023}. These fluxtubes push out disk material, powering short-term winds, thus influencing angular momentum transport \cite{Chatterjee2022}. (\textit{Right}) Time evolution of $\phi_{\rm BH}$ and $\eta$ for the $a=0.94$ simulation from Fig.~\ref{fig:eta}. We also indicate the time-averaged values from Fig.~\ref{fig:eta} with dashed lines. Flux eruptions repeatedly take away magnetic flux from the BH magnetosphere when magnetic forces overcome the accreting gas.
}
\label{fig:eruptions}
\end{figure}

Numerical simulations show that the BH magnetosphere can continue accumulating magnetic flux well past the time-averaged $\phi_{\rm BH}$ shown in Fig.~\ref{fig:eta}, slowly forming a current sheet layer, typically in the equatorial plane. At the point of magnetic flux over-saturation, magnetic reconnection occurs close to the BH, and nearly-cylindrical fluxtubes filled with vertical magnetic fields are ejected out into the disk. Vital questions about the ejection mechanism remain due to the complex 3D evolution of the reconnection event, such as: 
\begin{svgraybox}
\textit{(1) Is the formation of the current sheet only due to reconnection of the azimuthal magnetic field? (2) Does reconnection of the radial magnetic field trigger the eruption event, or does the azimuthal field also play a role? (3) What determines the limited azimuthal range covered by the expelled fluxtubes? (4) When are fluxtubes ejected and at what frequency in time? (5) Does the disk structure and dynamics also affect the ejection mechanism apart from shearing out the fluxtube once ejected into the disk?}
\end{svgraybox}

Figure~\ref{fig:eruptions} shows an example of how an ejected fluxtube propagates through the accretion disk in a GRMHD simulation. Additionally we show the time-evolution of the magnetic flux $\phi_{\rm BH}$ and the outflow efficiency $\eta$. We clearly see temporal oscillations in the time-series plots where there is a slow rise as magnetic flux is advected along with the gas towards the BH, thereby strengthening the outflow, and a fast decay due to fluxtube ejections. High resolution ideal and resistive GRMHD as well as GR-Particle-in-Cell simulations are actively investigating the ejection mechanism with a particular focus on the decay rate of the magnetic flux and its relation to the local reconnection rate \cite{Bransgrove2021, Ripperda2022, Galishnikova2023}. Further, once ejected, magnetic flux eruptions push against gas in the disk and can power disk winds, effectively regulating the disk angular momentum \cite{Chatterjee2022, Manikantan2024}, possibly alongside Maxwell stresses due to large-scale poloidal fields \cite{Narayan2012}. In summary, recent advances in simulations present an exciting opportunity to study magnetic flux eruptions in great detail, which could lead to new insights in the triggering, frequency and energetics of flares in accreting BHs (e.g., see Fig.~\ref{fig:flares}).

\section{Jet - ISM collisions in MAD accretion flows}
\label{sec:tilted}

\begin{figure}[t]
    \centering

    \includegraphics[width=\columnwidth,trim= 0pt 140pt 0pt 330pt, clip]{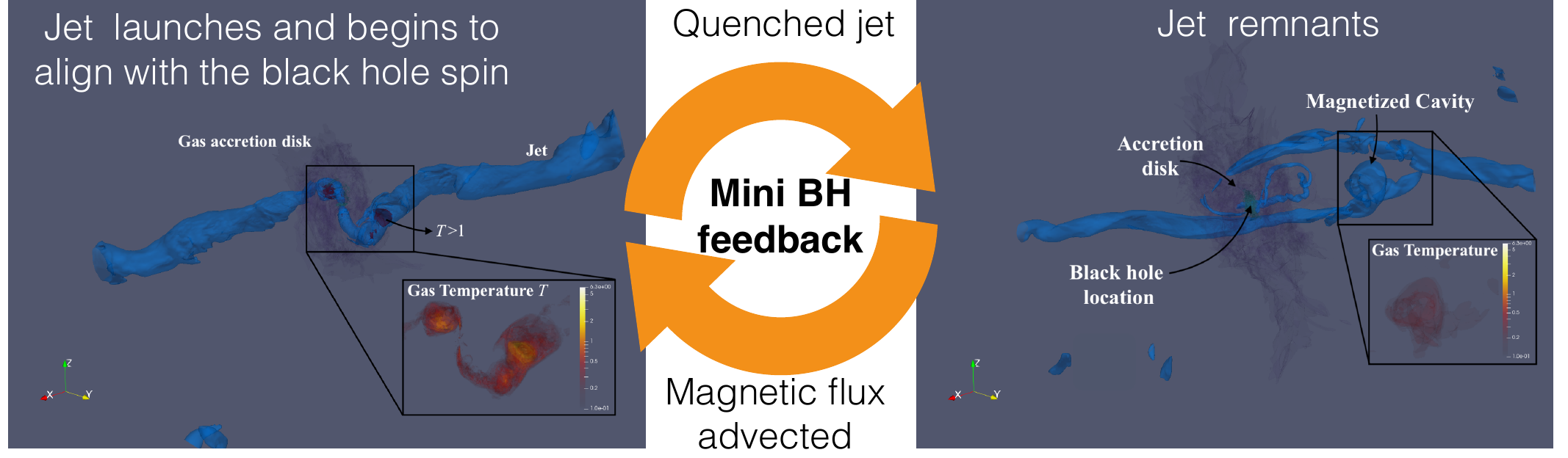}
    \includegraphics[width=\columnwidth,trim= 0pt 220pt 0pt 180pt, clip]{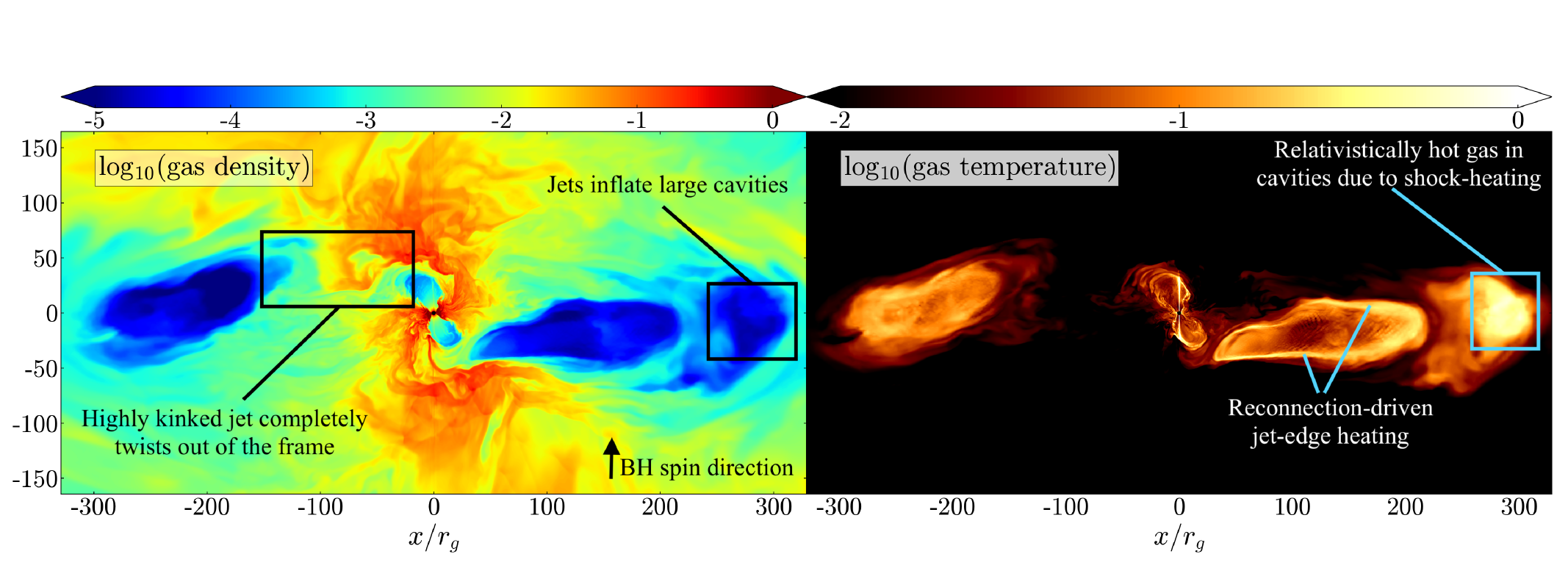}
    \caption{(\textit{Top}) Chatterjee et al. \cite{Chatterjee2023tiltedMADs} finds that a small-scale BH feedback loop can form for highly misaligned MAD accretion flows. As the newly formed jet strengthens over time due to the accumulation of magnetic flux at the horizon. When the jet begins to align with the BH spin, it collides with the accreting gas, disrupting the supply of magnetic flux to the horizon, leading to jet quenching. Once the jet subsides, new magnetic flux can be advected and brought to the BH, thereby restarting jet launching, and completing the feedback loop. In this particular simulation, the timescale for the feedback loop is around $5\times10^4 t_{\rm g}$, or around 3 days to 8 years for a $10^{6-9} M_{\odot}$ BH, which presents an interesting possibility for flaring states in radio-quiet AGN and quiescent X-ray Binaries. (\textit{Bottom left}) Vertical cross-section plot of gas density plot of the same simulation as above showing an extremely warped jet (blue) launched out of a spinning BH which is accreting from a $90^{\circ}$ misaligned disk (red). (\textit{Bottom right}) The distorted jet inflates shocked cavities in the ISM. The shocked gas shows relativistic temperatures ($T>1$ in relativistic units) suggesting significant non-thermal X-ray flaring. The jet also shows limb-brightened features as magnetic fields become squeezed between the jet and the ISM, which triggers reconnection and heats the gas. See this link for a movie of this simulation: \href{https://youtu.be/mTLbZgd5CtY}{https://youtu.be/mTLbZgd5CtY}.
    }
    \label{fig:feedback}
\end{figure}

Jets travelling through dense ISM environments, such as that found in quasi-spherical accretion flows, could be subject to large-scale kink instabilities, resulting in jet collapse. Gas infall in the polar regions of the BH could also inhibit jet production. Thus jet propagation is intimately connected to both the intrinsic jet power via the BZ process as well as environment properties. Powerful jets can burrow through the environment, creating a cocoon of magnetised gas that forms backflows and helps in jet collimation \cite{Rohoza2024}. For jets with smaller $\phi_{\rm BH}$, the jets can quickly lose stability and collapse, at times creating shocked cavities. Recent advances in computational resources have enabled GRMHD simulators to expand to larger length-scales and experiment with different types of accretion flows in order to self-consistently capture the complex dynamics of jet-ISM interactions \cite{Ressler2020, Lalakos2022, Cho2023, Kaaz2023, Ressler2023, Rohoza2024}.

Figure~\ref{fig:feedback} shows an interesting case when the incoming gas and magnetic fields are so misaligned from the BH spin vector such that the jet in the MAD state collides directly with the disk bulk, turns almost $90^{\circ}$ degrees and eventually falls apart \cite{Chatterjee2023tiltedMADs}. In this scenario, the jets actively regulate the advection of magnetic flux onto the BH horizon in a manner similar to flux eruptions (but with a lot more power!), creating a feedback loop as a result. The figure shows a plethora of interesting features such as jet remnants, magnetised filaments, cavities during the ``quenched jet phase'' and a twisted or bent jet during the ``jetted MAD phase''. The gas temperature ($T$) plot in Fig.~\ref{fig:feedback} shows that such systems could exhibit relativistically-hot gas that can produce X-ray emission alongside small-scale radio jets. We require more in-depth analysis of the radiative emission of these systems to see if such feedback loops can explain radio and X-ray flaring in radio-quiet AGN as an alternative mechanism to binary BH-disk interactions. See \cite{Chatterjee2023tiltedMADs} for details on the simulations and jet/disk evolution for such systems.

\section{Outlook}
\label{sec:outlook}

\begin{figure}[t]
    \centering

    \includegraphics[width=0.8\columnwidth,trim= 100pt 0pt 100pt 0pt, clip]{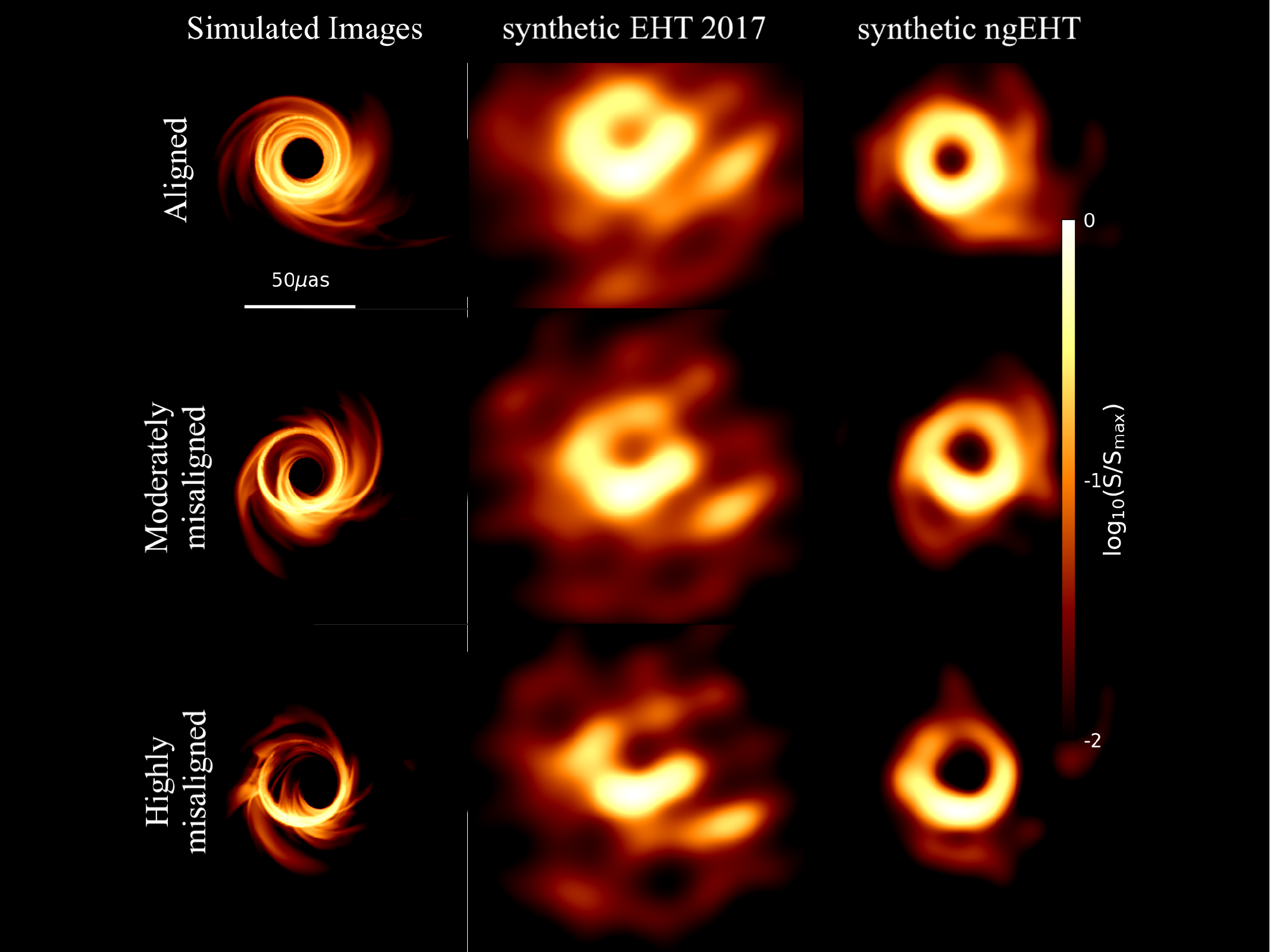}    
    \caption{Misaligned accretion onto spinning BHs result in bent jets that do not follow the BH spin direction. Misalignment significantly affects the evolution of the horizon-scale structure in these systems, altering the BH shadow size and shape as well as the extended jet structure. We show a snapshot of raytraced images for disks with misalignment angles of $0^{\circ}$, $30^{\circ}$ and $60^{\circ}$ \cite{Chatterjee2020}, measured at large distances, and their corresponding synthetic reconstructions at 230GHz (using the \texttt{eht-imaging} toolkit \cite{Chael2019ehtim}) with the 2017 EHT array \cite{EHTpaperIM872019} and the future ngEHT array \cite{Raymond2021, Johnson2023}, assuming the M87 BH mass and distance. The 2017 EHT array fails to properly capture the differences between these disks, demonstrating the need to resolve the disk-jet connection region by including more short-baseline arrays.
    }
    \label{fig:tilted_images}  
\end{figure}

Magnetically arrested disks present a uniquely rich laboratory for exploring magnetised plasmas in extreme conditions, from the micro-scale physics of reconnection, hotspots and shocks, to the meso-scales of BH accretion, disk warping in misaligned inflows and magnetic flux eruptions, and all the way to the large-scales of twisted and kink-unstable jets, magnetised backflows and cocoons. Figure~\ref{fig:tilted_images} shows one such example. We show raytraced horizon-scale images of 3 misaligned near-MAD models at sub-millimetre wavelengths (using the GR-radiative transfer code \texttt{BHOSS} \cite{Younsi2020}) and their reconstructions assuming the telescope array properties of the 2017 EHT array \cite{EHTpaperIM872019}, and a future large interferometric array (the next-generation EHT, or ngEHT \cite{Raymond2021, Johnson2023}), including only thermal noise for simplicity. We find that even after accounting for the systematic errors in imaging, it is possible to capture 3 primary features with the ngEHT array that indicate misaligned accretion: (1) the shape of the BH shadow region (i.e., the brightness depression in the center), given our knowledge of the source inclination, (2) the ring asymmetry, and (3) the warped disk-jet just outside the ring. We note that while these reconstructions are quite simplistic and more careful work needs to be done (much like that seen in the ngEHT Analysis Challenge papers; \cite{Roelofs2023, Chatterjee2023ngEHT}), it is encouraging that observing the horizon-scale accretion flow with such high resolution is possible either with ground-based arrays and potentially future space missions. 

Capturing the tell-tale signs of magnetic flux eruptions would be crucial to identify the MAD state. The equatorial plane reconnection event that triggers the eruption as well as the disk-fluxtube shearing/RT/KH instabilities could produce high-energy flares \cite{Ripperda2022}, while the expulsion of accretion material from the BH magnetosphere could produce a dip in the submillimetre flux \cite{Gelles2022} in AGNs like M87 and Sgr A* in addition to affecting polarimetric features in the horizon-scale image \cite{Jia2023} as well as the jet boundary \cite{Davelaar2023}. Further, the development of the MAD state in misaligned accretion often forces the inner part of the disk to ``magneto-spin align'' with the BH spin \cite{McKinney2013, Chatterjee2023tiltedMADs}, which have crucial consequences for interpreting the hard state of X-ray Binaries \cite{Fragile2023}, where some sources show quasi-periodic oscillations in the X-ray lightcurve. It is possible that flux eruptions may explain some quasi-periodic activity such as in AGNs \cite{Cui2023}, but power-spectral-density calculations of the MAD $\dot{M}$ time-series have not yet revealed promising results, primarily showing red noise. Evidently further investigations of the MAD state in standard torii need to be done, such as the dependence of accretion properties on simulation grid resolutions \cite{White2019, SosapantaSalas2024}, similar to analysis done for weakly-magnetised, MRI-dominated accretion flows \cite{Porth2019}. Similarly, understanding the ejection mechanism of fluxtubes holds important consequences for the interpretation of high-energy flares. Guided by the plethora of multi-wavelength observations of accreting BHs and jets, the future looks bright for numerical simulations dealing with accreting BHs with plenty of exciting scientific achievements to look forward to.

\begin{acknowledgement}
KC thanks Alex Raymond and Daniel Palumbo for their help with scripts to generate and reconstruct synthetic data from GRMHD simulations using \texttt{eht-imaging}. KC and RN were supported in part by grants from the Gordon and Betty Moore Foundation and the John Templeton Foundation to the Black Hole Initiative at Harvard University, and by NSF award OISE-1743747. ML was supported by the John Harvard, ITC and NASA Hubble Fellowship Program fellowships. AT acknowledges support from the NSF AST-2009884 and NASA 80NSSC21K1746 grants. AT was supported by BSF grant 2020747 and NSF grants AST-2107839, AST-1815304, AST-1911080, AST-2206471, OAC-2031997.This research was enabled by support provided by an INCITE program award PHY129, using resources from the Oak Ridge Leadership Computing Facility, Summit, which is a US Department of Energy office of Science User Facility supported under contract DE-AC05- 00OR22725, as well as Calcul Quebec (http://www.calculquebec.ca) and Compute Canada (http://www.computecanada.ca). This work has made use of NASA's Astrophysics Data System (ADS).
\end{acknowledgement}

\input{refs}

\end{document}

%% file: refs.tex
%%%%%%%%%%%%%%%%%%%%%%%% referenc.tex %%%%%%%%%%%%%%%%%%%%%%%%%%%%%%
% sample references
% %
% Use this file as a template for your own input.
%
%%%%%%%%%%%%%%%%%%%%%%%% Springer-Verlag %%%%%%%%%%%%%%%%%%%%%%%%%%
%
% BibTeX users please use
% \bibliographystyle{}
% \bibliography{}
%
%\biblstarthook{